\documentclass{PoS}
\usepackage{latexsym,amsmath,amstext}
\usepackage{mathrsfs}
\usepackage[utf8x]{inputenc}
\usepackage{epsfig}
\usepackage{float}
\usepackage{slashed}
\usepackage{amssymb,fontenc,times,mathptmx,graphicx}
\title{Renormalized quasi parton distribution function of pion}

\author{
\speaker{Nikhil Karthik}$^1$, Taku Izubuchi$^{1,2}$, Luchang Jin$^{2,3}$, Christos Kallidonis$^{4}$, 
Swagato Mukherjee$^{1}$, Peter Petreczky$^{1}$, Charles Shugert$^{1,4}$, Sergey Syritsyn$^{2,4}$\\
\llap{$^1$}Physics Department, Brookhaven National Laboratory, Upton, NY 11973, USA \\
\llap{$^2$}RIKEN-BNL Research Center, Brookhaven National Lab, Upton, NY, 11973, USA \\
\llap{$^3$}Physics Department, University of Connecticut, Storrs, Connecticut 06269-3046, USA \\
\llap{$^4$}Department of Physics and Astronomy, Stony Brook University, Stony Brook, NY 11794, USA}

\abstract{ 
We present preliminary numerical results on the connected piece of
the quasi-PDF of pion as determined using Wilson-Clover valence
fermions on HISQ ensembles. We discuss its non-perturbative
renormalization in RI/MOM scheme with and without removal of the
divergent self-energy part, and compare its running with expectation
from perturbation theory. We also discuss the matching of pion QPDF
to PDF, and various systematic effects associated with it.
}

\FullConference{The 36th Annual International Symposium on Lattice Field Theory - LATTICE2018\\
                22-28 July, 2018\\
                Michigan State University, East Lansing, Michigan, USA.}
\ShortTitle{qpdf of pion}
\begin{document}

\section{Introduction}

For hard scattering processes involving hardons, such as the deep
inelastic scattering of leptons on hadrons, the total cross-section
becomes a convolution of the perturbatively calculable partonic
cross-section and the universal, non-perturbative parton distribution
function~\cite{Feynman:1969ej}.  In the light-cone gauge, the parton
distribution function has the familiar interpretation of the
probability of finding a parton with a fraction $x$ of the energy
of a fast moving hadron. Until recently, the PDFs were obtained
through global fits to the experimental
data~\cite{Martin:2009iq,Nadolsky:2008zw,Ball:2008by,Aaron:2009aa}.
Naively, one cannot use the standard lattice Monte Carlo to determine
the PDFs since it involves quark and anti-quark operators separated
along the light-cone, which therefore has a sign problem associated
with it. Recently, the computation of PDFs using quasi-PDF, which
involves equal-time Euclidean correlations of spatially separated
quark-antiquark pair, of fast moving hadron and then matching to
PDF using large momentum effective theory approach has been proposed
to be a solution~\cite{Ji:2013dva,Ji:2014gla}.  It is the aim of
this Lattice proceeding to present our preliminary determination
of the PDF of pion using the qPDF approach as well as to discuss
various systematic effects associated with this procedure.  Recently,
a global fit analysis of pion PDF was obtained in~\cite{Barry:2018ort}.
The reader can refer to~\cite{Chen:2018fwa} for the qPDF determination
of valence PDF of pion using different lattice spacing, ensemble
and matching procedure than what is used here.

Below, we outline the steps involved in the quasi-PDF approach and
note the different places which can potentially lead to systematic
effects that need to be controlled in the extraction of PDF:
\begin{enumerate}
\item First, one computes the bare quasi-PDF using the three-point to two-point
function ratio given by
\begin{equation}
q_\Gamma(z,P_z;\Delta t) = \frac{\langle \hat\pi^\dagger_S(P_z,\Delta t) O_\Gamma(z) \hat\pi_S(P_z,0)\rangle}{\langle \hat\pi_S^\dagger(P_z,\Delta t) \hat\pi_S(P_z, 0)\rangle},
\end{equation}
where $\hat\pi_S(P_z,t)$ is the 
pion operator inserted at time-slice $t$
with the spatial momentum $(0,0,P_z)$ and
smeared using a method $S$,
$\Delta t$ is the source-sink separation and the 
bilocal qPDF operator with Dirac structure $\Gamma$ is given by
\begin{equation}
O_\Gamma(z;\tau) = \overline{u}(x)\Gamma W_{x,x+z \hat z} u(x+z \hat z),
\end{equation}
where $W$ is the Wilson from $x$ to $x+z \hat z$.
There is both a connected and a disconnected piece to the above
matrix element. Here, we only focus on the connected part of this.
The presence of finite $\Delta t$ is one systematic effect in the
calculation. This is discussed in more detail in the accompanying
Lattice proceeding.

\item Then, one renormalizes the bare quasi-PDF using nonperturbative renormalization (NPR) procedure. Here, we use the 
RI-MOM renormalization condition with $\slashed{p}$ projection~\cite{Chen:2018xof}
\begin{equation}
q^R_{\gamma_t}(z,P_z;p^R,\Delta t)=Z_{\gamma_t\gamma_t}(z,p^R) q_{\gamma_t}(z,P_z;\Delta t)
\end{equation}
with the renormalization constant $Z(z,p^R)$ determined using the following condition imposed on 
the amputated quark qPDF, $\Lambda(z,p)$ using external quark states at momentum $p=p^R$:
\begin{eqnarray}
{\rm Tr}\left(\slashed{p}_R \Lambda(z,p_R)\right)  = 12 p^R_t e^{-i p^R_z z}.
\end{eqnarray}
In the case of $\Gamma=\gamma_z$, one should also take care of
mixing between the bare $\Gamma=\gamma_z$ and $\Gamma=1$
qPDFs~\cite{Constantinou:2017sej}. The running of the renormalized
quark qPDF is simply the dependence of the renormalized amputated
quark qPDF $\Lambda(z,p; p^R)$ on $p$, with $p$ slightly away from
$p^R$. Any mismatch between the nonperturbative and 1-loop perturbative
running could be another source of systematic error.

\item The renormalized real space qPDF $q^R_{\gamma_t}(z,P_z;p^R,\Delta t)$ is Fourier transformed 
to $\tilde{q}_{\gamma_t}(x,P_z;p^R,\Delta t)$, where $x$ is the momentum fraction of $P_z$. That is
\begin{equation}
\tilde{q}_{\gamma_t}(x,P_z;p^R,\Delta t) = 2E(P_z)\int_{-\infty}^{\infty}\frac{dz}{4\pi} q^R_{\gamma_t}(z,P_z;p^R,\Delta t) e^{-i x P_z z},
\end{equation}
where $E(P_z)$ is the energy. In order to take the Fourier
transform, the data has to be interpolated by a continuous curve.
More importantly, one has actual data for the real space qPDF only
over a certain finite range of $z$. In this case, the absence of
data for larger $z$ and what one does to take care of this could
lead to further systematic uncertainties.

\item The final step is the matching of the Fourier transform of
the renormalized qPDF to PDF at a factorization scale
$\mu$~\cite{Ji:2013dva,Stewart:2017tvs}. For this, we convolute
$\tilde{q}(x)$ with the matching coefficient. 
In principle, the effect of this procedure should completely remove
any dependence on $p^R$ and transmute it to a dependence on $\mu$.

We present results using the a=0.06 fm, HISQ sea quark ensemble
generated by the HotQCD collaboration. On this, we use 1-HYP smeared
Wilson-clover valence quarks to determine the qPDF. More details
on the measurement techniques is given in another proceeding
accompanying this one.

\section{Effectiveness of perturbation theory to describe NPR}
\end{enumerate}
\begin{figure}
\center
\includegraphics[scale=0.45]{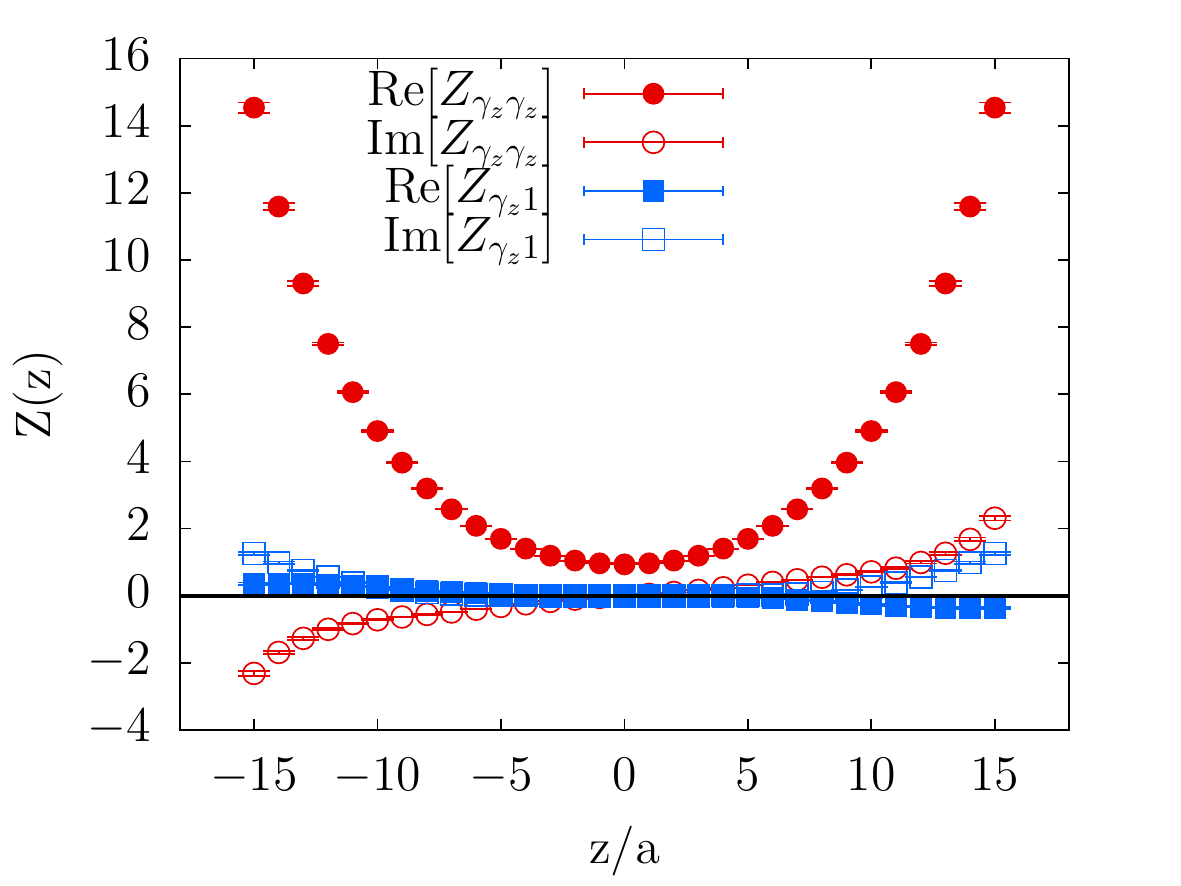}
\includegraphics[scale=0.45]{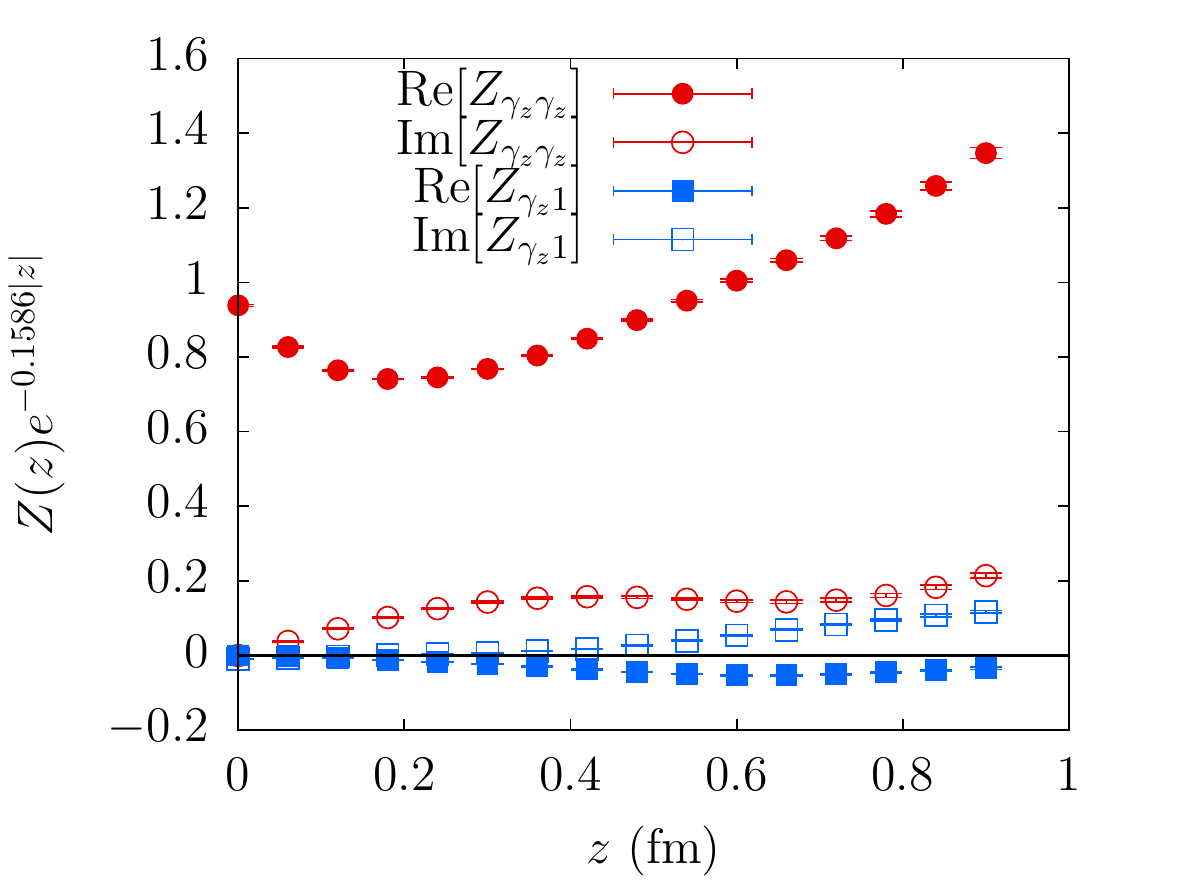}
\caption{Renormalization constants $Z_{\Gamma\Gamma'}$ including (left) and without including (right)  the 
linearly divergent self-energy contribution
from the Wilson line connecting the quark and anti-quark. In the two panels, the data for 
the diagonal $Z_{\gamma_z,\gamma_z}$ as well as the mixing with the scalar $Z_{\gamma_z,1}$ are shown.
The solid symbols are for their real parts while the open ones for their imaginary parts.
}
\label{showzfac}
\end{figure}

The only non-perturbative ingredient that is supposed to go into
the PDF computation is the projection to pion state itself, which
is taken care of by using the pion source and sink at large
separations. The other computations, including the renormalization
of the real space qPDF in the presumably relevant range of $|z| <
1$ fm and the subsequent matching should be perturbative. The
subtlety with regard to the renormalization constant is the presence
of divergent self-energy contribution, $e^{c |z|}$, in the bare
Wilson line $W_{x,x+z\hat z}$ connecting the quark and anti-quark.
This should not be problem for perturbative calculation since this
multiplicative factor should cancel between the NPR $Z$-factor and
the bare qPDF. However, we check that what remains after the
subtraction of the self-divergent piece is $O(1)$ as one would
expect from perturbative calculation.  For the one-level HYP smeared
link, we use the nonperturbatively determined value $c=0.1586$~\cite{Bazavov:2018wmo}.  In
Figure \ref{showzfac}, we show the the diagonal renormalization
factor $Z_{\gamma_z,\gamma_z}$ as well as the offdiagonal
$Z_{\gamma_t,\gamma_t}$ as a function of $z$. The red points are
the real parts while the blue ones are the imaginary parts. In the
left panel, we show the entire renormalization factor. In the right
panel, we show the renormalization factors after multiplying by
$e^{-c |z|}$. As one can see, the renormalization factor after the
subtraction of the divergent piece, $Z_{\gamma_z\gamma_z} e^{-c|z|}$,
is surprisingly close to one even for distances upto 1 fm. We see
a similar behavior for $Z_{\gamma_t\gamma_t}$ as well.

\begin{figure}
\center
\includegraphics[scale=0.5]{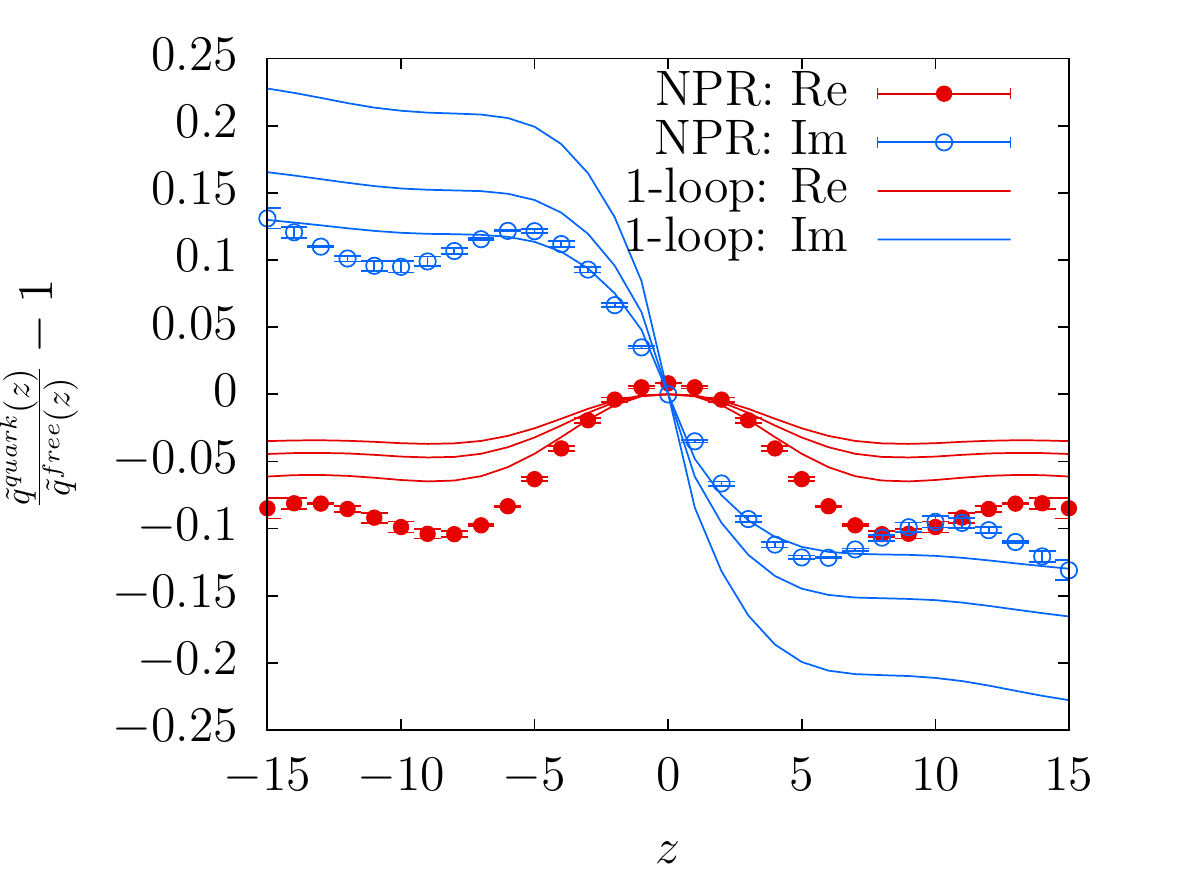}
\caption{Comparison of quark qPDF to perturbation theory. Refer text below for explanation.}
\label{qpert}
\end{figure}

In order to see if we see a quantitative agreement of the $Z$ factors with perturbative 
calculation, we look for the running of
\begin{equation}
\zeta(z;p)=\frac{{\rm Tr}(\slashed{p} \Lambda^R(p;p_R))}{{\rm Tr}(\slashed{p} \Lambda^R(p_R;p_R))}-1,
\end{equation}
where $\Lambda^R(p;p_R)$ is the amputated qPDF with momentum $p$ and 
renormalized at momentum $p^R$. 
The term ${\rm Tr}(\slashed{p} \Lambda^R(p_R;p_R))$ is the free field value by 
the renormalization condition. 
Since the corresponding expression for $\zeta$ from 
perturbation theory is simpler when $p^R_z=p_z$, we study this case here. 
In Figure \ref{qpert}, we show the real and imaginary parts of 
$\zeta(z;p)$ as a function of $z$ for $p=(0.86,0.86,1.61,0.86)$ GeV with the 
renormalization condition set at $p_R=(1.28,1.28,1.61,1.28)$ GeV. The three curves
are the corresponding 1-loop perturbative results with the strong coupling $\alpha_S$ determined 
at $2 p^R$, $p^R$ and $p^R/2$ respectively in order to quantify the uncertaintly in the scale
to be used in the 1-loop calculation. While the agreement between the data and the curves is good for the imaginary part for 
$|z|< 0.5$, only a qualitative agreement is seen in the case of the real part. However, this $p$ dependence
is only sub-leading in the case of the real part, and hence this disagreement might not be an issue given 
other uncertainties.

\section{qPDF and PDF of pion}
\begin{figure}
\center
\includegraphics[scale=0.65]{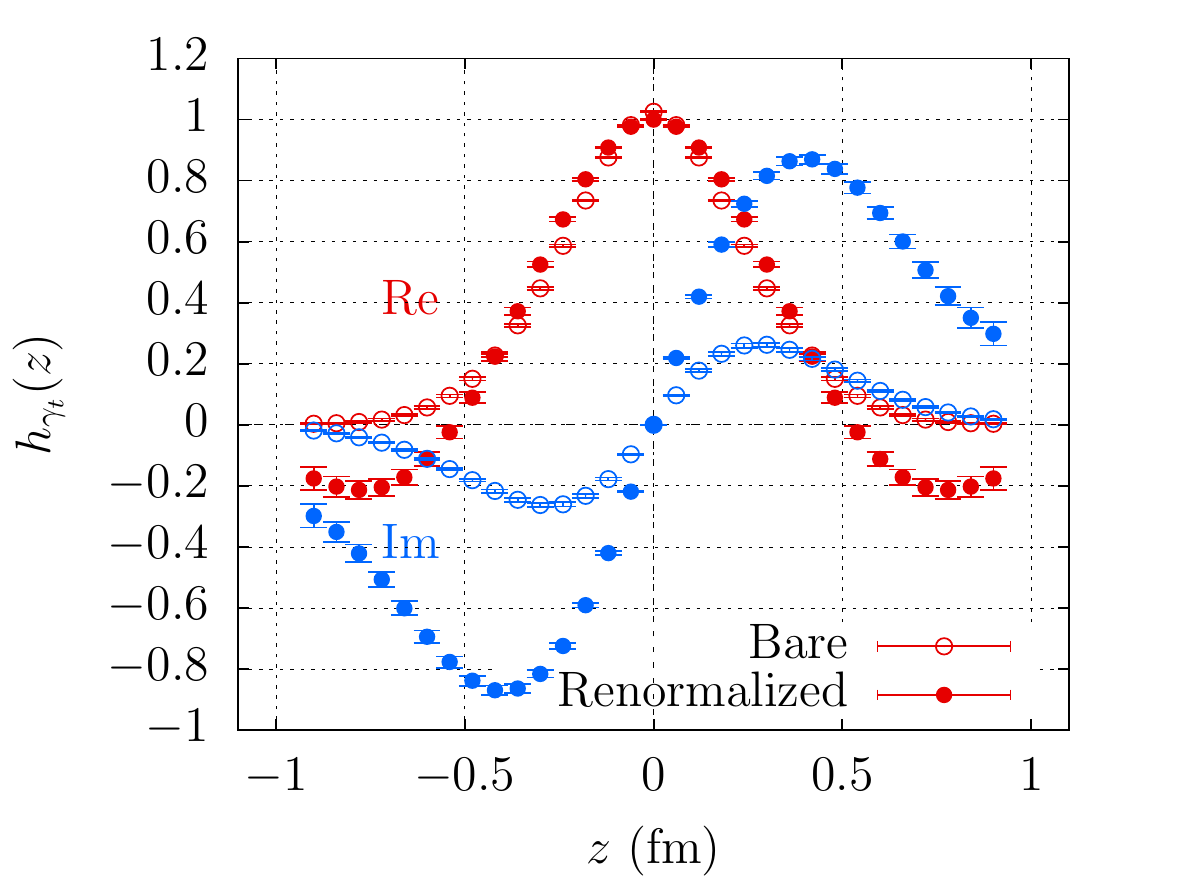}
\caption{The real space qPDF is shown as a function of $z$. The real and imaginary parts are shown using 
red and blue points respectively. The open symbols are for the bare qPDF and the solid ones are for the 
renormalized qPDF.}
\label{showqpdf}
\end{figure}

In Figure \ref{showqpdf}, we show the bare qPDF (open symbols) and
the corresponding renormalized qPDF (filled symbols) as a function
of $z$. The red points are the real parts and the blue ones are the
imaginary parts. The effect of renormalization is to lift the
exponentially suppressed bare qPDF at larger $z$. This effect is
more significant in the imaginary part than in the real part of
qPDF. We have data for qPDF from $z/a=-15$ to 15 in the above plot
which ranges from -1 fm to 1 fm in physical units.

\begin{figure}
\center
\includegraphics[scale=0.5]{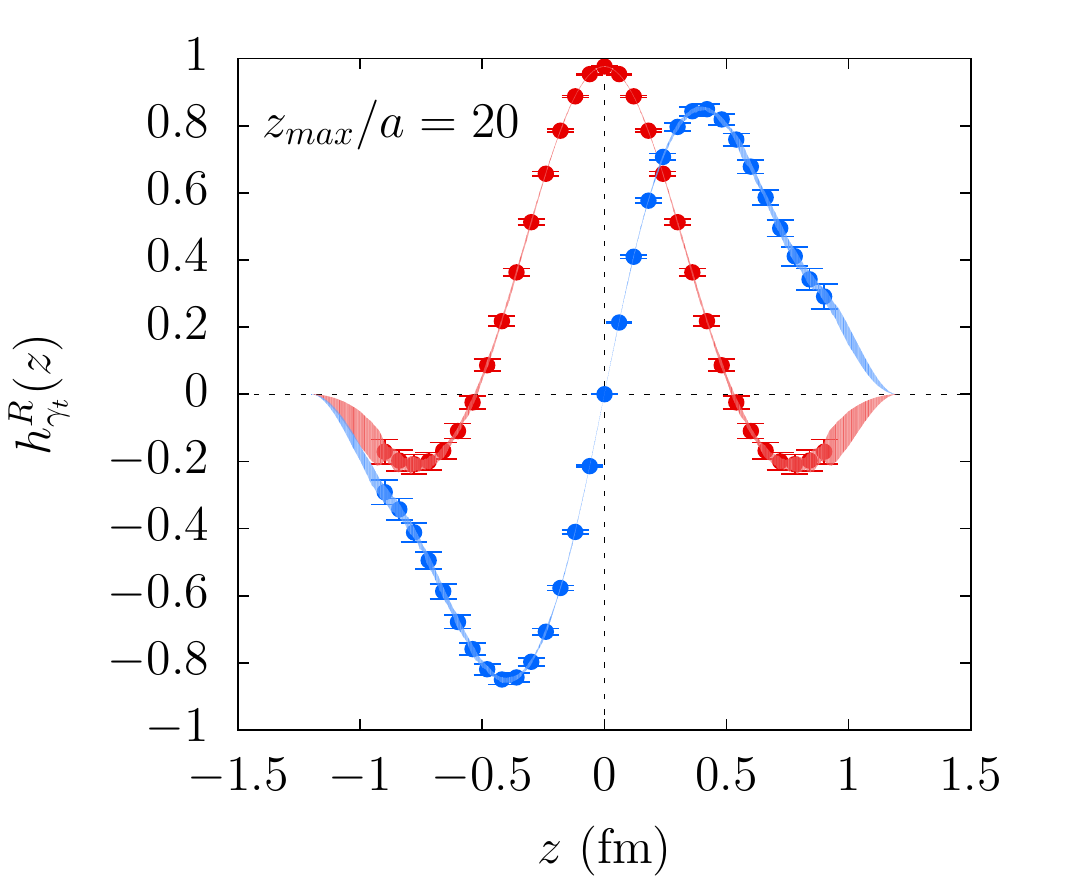}
\includegraphics[scale=0.5]{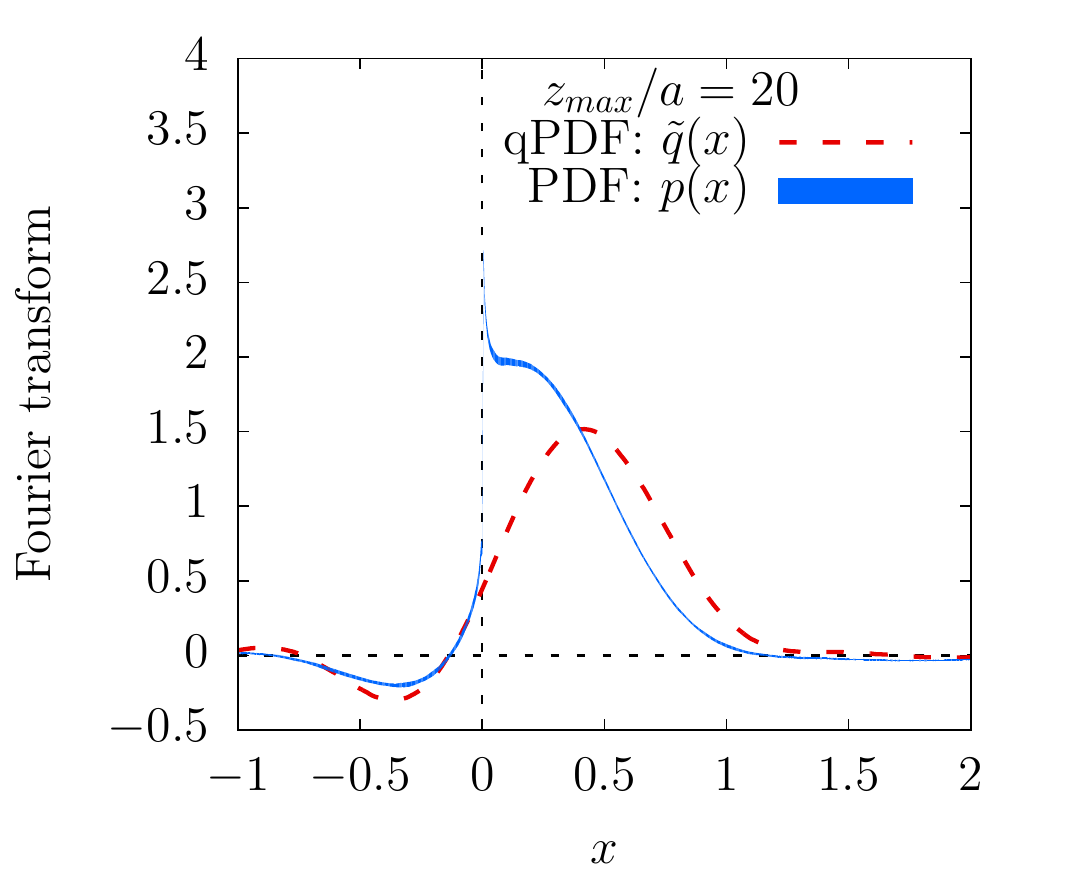}

\includegraphics[scale=0.5]{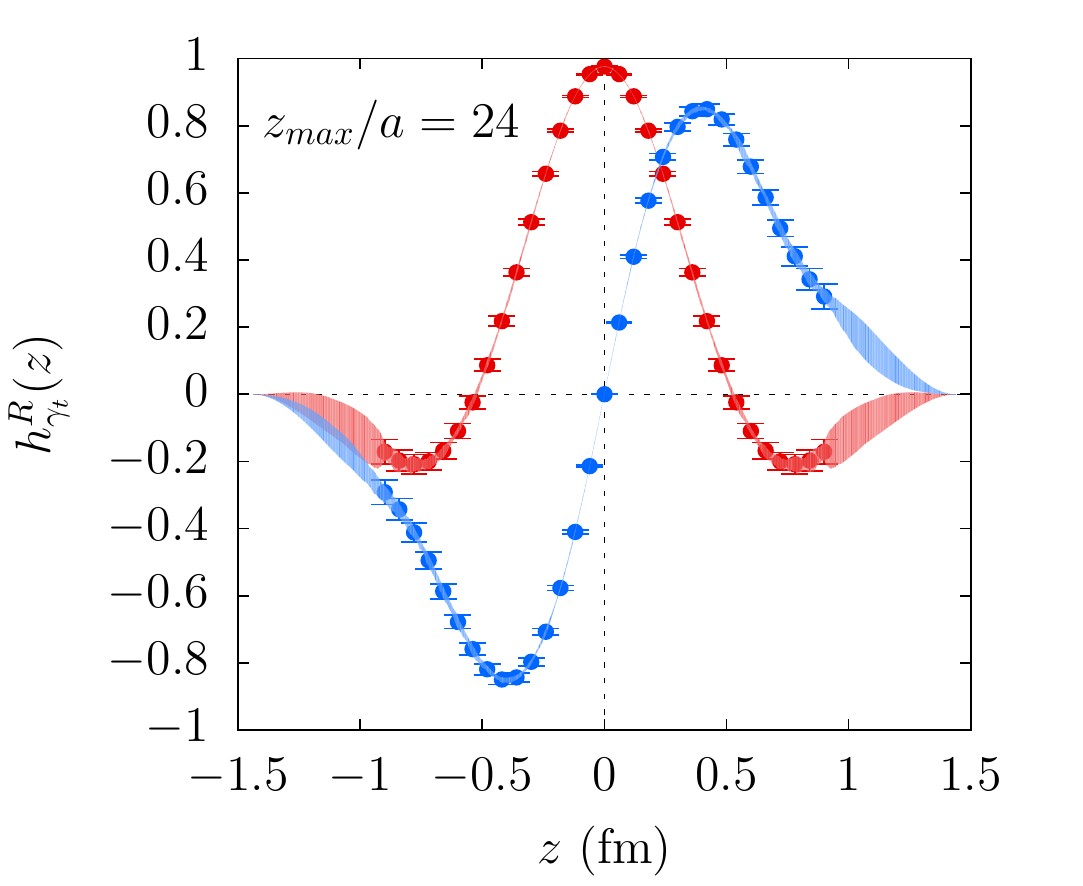}
\includegraphics[scale=0.5]{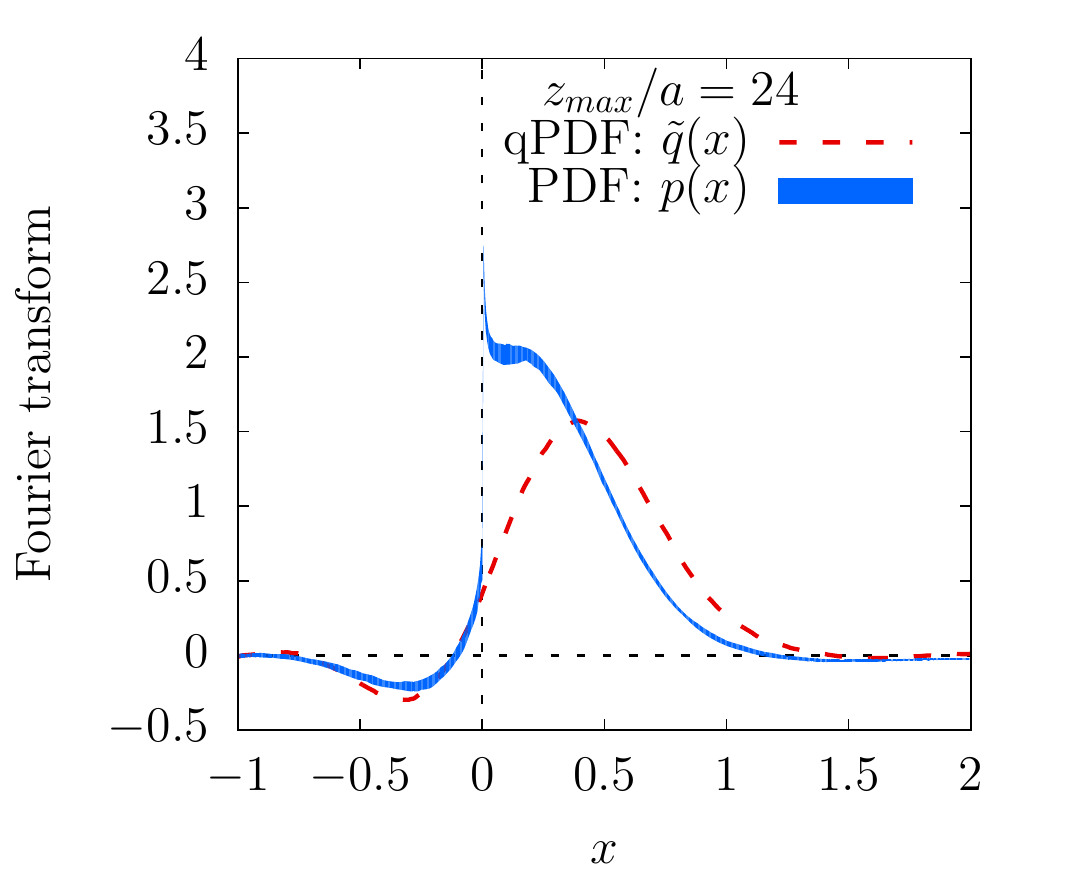}
\caption{Cubic interpolation and extrapolation to $z_{\rm max}/a=20$ (top-left) and 
$z_{\rm max}/a=24$ (bottom-left) are 
shown. The corresponding Fourier transform of qPDF and the matched PDF at $\mu=3.2$ GeV are shown in the 
right panels. The description of the plots is in the text below.}
\label{longdist}
\end{figure}

In order to take
the Fourier transform, we interpolate the real and imaginary parts
of $q^R(z)$ using cubic spline. To quantify the effect of $q^R(z)$
at larger $z$ where we do not have data currently, we extend the
spline to an arbitrary value of $|z|=z_{\rm max}$, at which point
we set the spline and its derivative to 0.  By varying the value
of $z_{\rm max}$, we quantify the effect of large distance part of
qPDF on the extracted PDF.  It is the expectation that the large
distance behavior of qPDF should not affect the short distance
physics of PDF. 
The top-left panel of Figure \ref{longdist} shows the real space real and imaginary parts of qPDF,
and the cubic interpolation (bands) which is extended upto $z_{\rm max}/a=20$. The right 
panel shows the Fourier transform of the interpolation shown on the left, using dashed 
lines. The PDF obtained by applying the matching formula to the 'dashed lines' is shown 
using the blue 1-$\sigma$ error bands. The corresponding plots for $z_{\rm max}/a=24$ is shown in the 
bottom panels. The effect of the extrapolation is mainly to increase the error in $\tilde{q}(x)$ and 
matched PDF $p(x)$ without affecting the central values. This is reassuring that the unknown long distance 
contribution of qPDF does not play a significant role. However, this could still be specific to the way we are 
taking care of the long distance behaviour of $q(z)$. It needs to be seen if this independence of long distance 
physics is seen with other models of extrapolation to larger $|z|$.

\begin{figure}
\center
\includegraphics[scale=0.6]{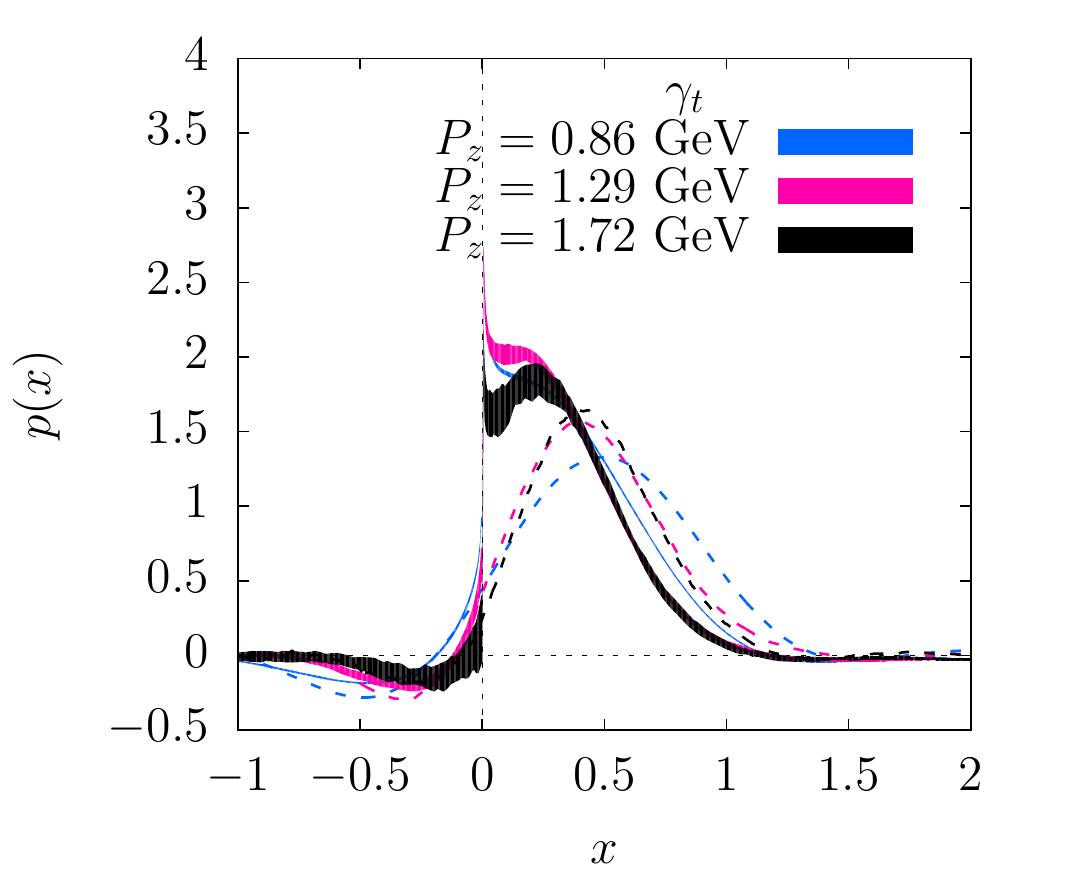}
\includegraphics[scale=0.6]{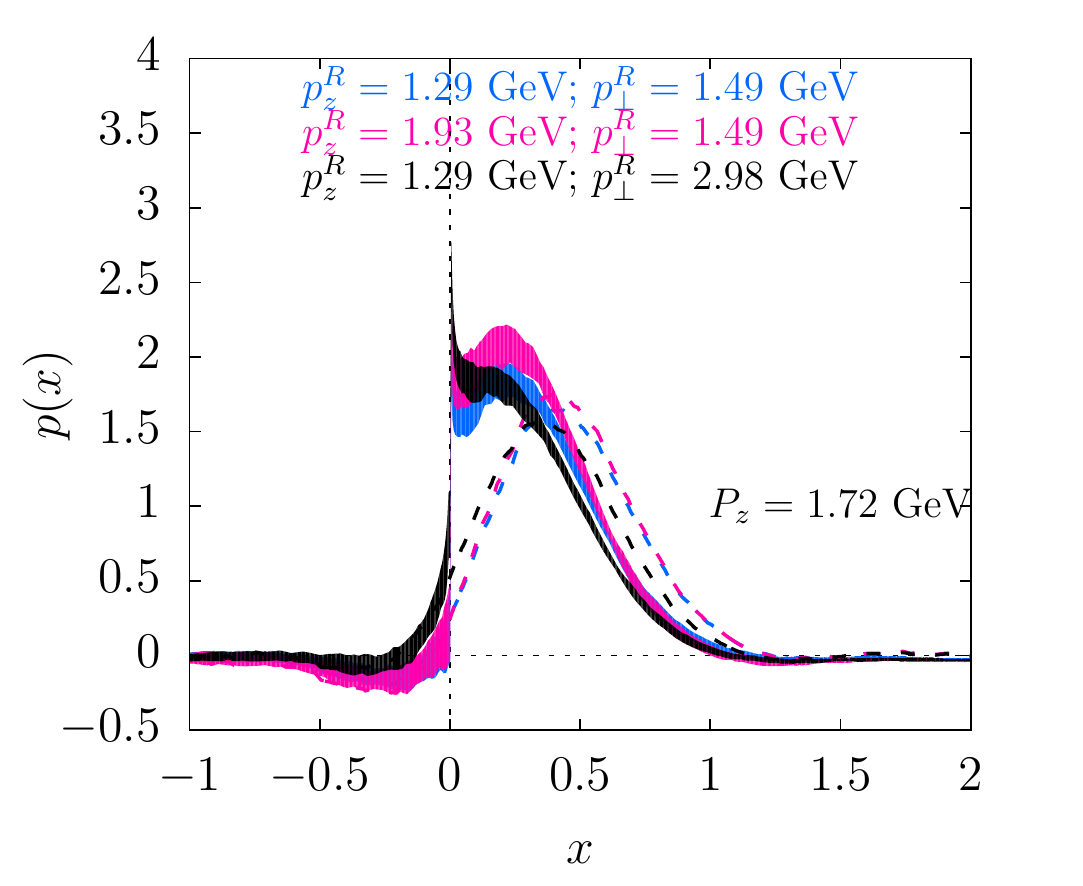} 

\includegraphics[scale=0.6]{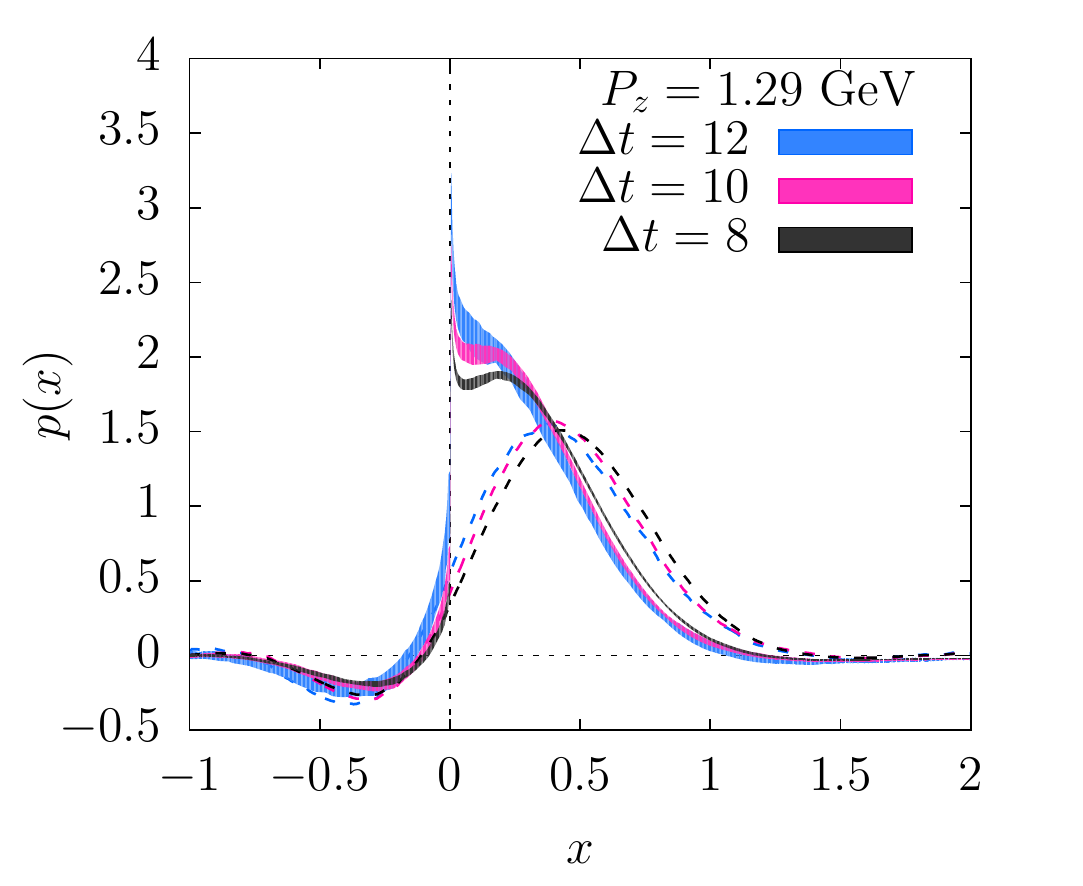}
\caption{Matched valence PDF of pion. In all the panels, the bands are 
the $1-\sigma$ error bands for the PDF while the dashed curves of the same color 
are the corresponding $\tilde{q}_{\gamma_t}(x)$ used to obtain the PDF.  (Top left) The $P_z$ dependence of PDF is shown.
(Top right) The dependence of PDF on the renormalization scale for the qPDF is shown.
(Bottom) The dependence on source-sink separation $\Delta t$ is shown.}
\label{pdfsystematics}
\end{figure}

Now we focus on the matched PDF that we obtain at a scale $\mu=3.2$ GeV using 
perturbative matching with the strong coupling $\alpha_S(\mu)=0.234$.  The first systematic
is the approach to $P_z\to\infty$ limit. We use $P_z=$0.86 GeV,
1.29 GeV and 1.72 GeV, which are smaller compared to the lattice
spacing of 3.28 GeV and much larger compared to the pion mass of 0.3
GeV. In the left panel of Figure \ref{pdfsystematics}, we show
$\tilde{q}\left(x,P_z;p^R_z=1.28 {\rm GeV}, p^R_\perp=2.21 {\rm GeV}\right)$
and the matched PDF $p\left(x,\mu=3.2{\rm GeV}\right)$, at different $P_z$
with every other parameter held fixed.
There seems to be convergence by looking at the consistency between the 
$P_z=1.29$ and 1.72 GeV results for the PDF. It is also 
reassuring that the matched PDF vanishes for $x > 1$. 
For $|x| < \Lambda_{\rm QCD}/P_z$, the long distance
non-perturbative effects would become important. For the largest momentum $\Lambda_{\rm QCD}/P_z \approx 0.1$.
In the right panel of Figure \ref{pdfsystematics}, 
we show the dependence of the matched PDF on the renormalization scale of the qPDF. The 
different bands correspond to different renormalization points $(p^R_z,p^R_\perp)$ as specified. The matching 
procedure should remove any $p^R$ dependence present in $\tilde{q}(x)$. A mild $p^R$ dependence 
is seen in the matched PDF data at the largest momentum consistent with the expectation. In the bottom
panel of Figure \ref{pdfsystematics}, we show the effect of source-sink separation $\Delta t$ on the matched PDF at 
fixed values of $P_z$ GeV, $p^R$ and $z_{\rm max}$. There is a tendency for the data in the range $x>1$ to move closer to zero
while the data in the range $0<x<0.5$ to increase as the source-sink separation is increased from 
8 lattice units to 12, thereby moving towards the
phenomenological expectation. 

\section{Conclusion}
We presented results on the non-perturbative renormalization constants and showed that they are of order 1 after  the
removal of the self energy divergence of the Wilson line 
even upto  quark anti-quark separations of 1 fm. But, on a quantitative level, 
a slight disagreement with the perturbative prediction of the running of the quark qPDF was seen. Perhaps, this slight
disagreement is not of consequence given other sources of uncertainties. Then, we presented the fourier transform 
of the connect part of the pion quasi PDF and its mild dependence on the behaviour of qPDF for distances greater than 1 fm. 
Finally, we presented the matched PDF of pion and discussed its dependence on the pion momentum, the renormalization 
scale of the qPDF and the source-sink separation.

The work was supported by the U.S. Department of Energy, Office of
Nuclear Physics through the Contract No. DE-SC001270, BNL LDRD
project No. 16-37,  and Scientific Discovery through Advance Computing
(ScIDAC) award ”Computing the Properties of Matter with Leadership
Computing Resources”.  SS also acknowledges support by the RHIC
Physics Fellow Program of the RIKEN BNL Research Center.  Computations
were carried out using USQCD resources at JLab and BNL. This research
also used an award of computer time provided by the INCITE program
at the Oak Ridge Leadership Computing Facility, which is a DOE
Office of Science User Facility supported under Contract
DE-AC05-00OR22725.

\bibliographystyle{JHEP}
\bibliography{biblio}
\end{document}